\documentclass[manuscript]{aastex}

\usepackage{graphicx,natbib}
\newcommand{\id}{ {\rm d} }   
\newcommand{\mps}{m\,s$^{-1}$}
\newcommand{\kgs}{kg\,s$^{-1}$}
\newcommand{\kmps}{km\,s$^{-1}$}

\shorttitle{Moat flow system around sunspots in shallow subsurface layers}
\shortauthors{\v{S}vanda et al.}

\begin{document}

\title{Moat flow system around sunspots in shallow subsurface layers}
\author{Michal \v{S}vanda}
\affil{Astronomical Institute, Academy of Sciences of the Czech Republic (v. v. i.), Fri\v{c}ova 298, CZ-25165 Ond\v{r}ejov, Czech Republic}
\affil{Astronomical Institute, Charles University in Prague, Faculty of Mathematics and Physics, V Hole\v{s}ovi\v{c}k\'ach 2, CZ-18000 Prague 8, Czech Republic}
\email{michal@astronomie.cz}
\author{Michal Sobotka}
\affil{Astronomical Institute, Academy of Sciences of the Czech Republic (v. v. i.), Fri\v{c}ova 298, CZ-25165 Ond\v{r}ejov, Czech Republic}
\author{Tom\'a\v{s} B\'arta}
\affil{Astronomical Institute, Charles University in Prague, Faculty of Mathematics and Physics, V Hole\v{s}ovi\v{c}k\'ach 2, CZ-18000 Prague 8, Czech Republic}

\begin{abstract}

We investigate subsurface moat flow system around symmetrical sunspots of McIntosh type H and compare it to the flow system within supergranular cells. Representatives of both types of flows are constructed by means of statistical averaging of flow maps obtained by time--distance helioseismic inversions. We find that moat flows around H-type sunspots replace the supergranular flows but there are two principal differences between the two phenomena: The moat flow is asymmetrical, probably due to proper motion of sunspots with respect to the local frame of rest, while the flow in the supergranular cell is highly symmetrical. Further, the whole moat is a downflow region, while the supergranule contains the upflow in the centre, which turns into the downflow at some 60 \% of the cell radius from its centre.
We estimate that the mass downflow rate in the moat region is at least two times larger than the mass circulation rate within the supergranular cell.

\end{abstract}

\keywords{Convection --- Sunspots --- Sun: helioseismology}

\section{Flows around sunspots}
Sunspots are probably the most intensively studied topic of solar physics. The strong magnetic field, which is responsible for the sunspot appearance and evolution, significantly affects the pattern of convection and plasma flows in the upper layers of the solar convection zone. Sunspots mainly suppress an upward propagation of the heated plasma in their cores (umbrae) and harbour strong radial outflows in penumbrae, usually termed the Evershed flow \citep{1909MNRAS..69..454E}. 
The umbra is in principle stationary. Deep in its photosphere, localised small-scale (150 km) upflows of 0.6--1.4~\kmps{} are observed inside most of umbral dots and downflows of 0.4--1.9~\kmps{} at the edges of some of them \citep{2010ApJ...713.1282O}. The penumbra at the photospheric level is dominated by the horizontal Evershed flow with a magnitude of 2--5~\kmps. Vertical upflows of about 1~\kmps{} are observed in the inner and downflows of 0.5--1~\kmps{} in the outer penumbra. At the sunspot border, the downflows may reach 5~\kmps. All these flows appear in strongly localised patches, including the well-known filamentary structure of the Evershed flow, and change with time. On azimuthal average, the flow is essentially horizontal with a small upward component (200--300~\mps) in the inner and a small downward component (400~\mps) in the outer penumbra \citep{2011PhDT.......137F}.

Eventhough not directly seen, penumbrae of evolved sunspots are usually surrounded by an additional outflow region, a moat \citep{1969SoPh....9..347S}. This intriguing region around sunspots seems to be present mostly around evolved and decaying spots and plays a role in transporting flux away from the spot, hence contributing to its decay. \cite{1973SoPh...28...61H} showed that the magnetic elements carried by the moat flow have both polarities, however the net flux appears to have the same sign as the parent spot. The total flux transported from the spot seems to correspond to the decayed magnetic flux of the spot. 

The amplitude of the moat flow is usually of about 500~\mps. The moat seems to be present only on the side of the spots where the penumbra exists \citep{2008ApJ...679..900V}. \cite{Deng2007}, however, reported a case of a persisiting moat flow after the penumbra decayed. The width of the moat-flow region depends only weakly on the size of the parent spot \citep{2007AA...472..277S}. The latter authors analysed motions of photospheric granules in vicinity of sunspots and found a significant asymmetry of moats. Their areas are deformed in the east-west direction in such a way that the western part is narrower and the eastern one is broader than the average moat width. Also flow velocities are asymmetric, lower by 10~\% in the western part than in the eastern one. \cite{2007AA...472..277S} explained this asymmetry by an interaction of the moat outflow with the measured westward motion of sunspots (about 100~\mps) in the local frame of rest. As the sunspot flux tube moves through the convection zone, the subsurface flows around it are deformed due to a gas viscosity, similarly to a wake behind a sailing ship. There also seem to exist single cases of sunspots where the moat flow is purely symmetric \citep{BM2010}. The connection between the Evershed flow and the moat flow was studied by several authors with varying results. The modern studies \citep[such as][]{2013AA...551A.105L} seem to slightly prefer a physical origin of the moat flow to be distinct from the origin of the Evershed flow.

Attempts to explain the observed properties of the moat flow were seen in literature already shortly after the discovery, when the observers pointed at the resemblance of the moat flow and supergranular flows.  Supergranulation is a mid-scale convection-like velocity pattern with cells having a typical size of $\sim30$~Mm and a lifetime of around one day \citep[e.g.][]{2008SoPh..251..417H,Roudier2014} covering all the solar surface. Indeed, the predominantly horizontal outflow velocity field within supergranules with a peak flow of around 500~\mps{} reminds the outflow in the moat. Studies showed that the moat flow was usually faster than an average supergranular flow and moats also lived longer (for several days) than the ordinary supergranules. So it was \cite{1974MNRAS.169...35M}, who first suggested that the moat flow was essentially a supergranular flow with a sunspot in the middle of the cell. 

The flow system around sunspots was a target also for modellers. The 2-D simulations by \cite{2000MNRAS.314..793H} showed a model, where the flux tube responsible for the confinement of the sunspot umbra is first surrounded by an inflow region, which in the case of an evolved sunspot is hidden below the penumbra and hence not seen by surface observations. Further away a large outflow region appears, observable as the moat flow in the surface layers. The moat-like flow is also present in the state-of-the-art simulations, such as those of \cite{2009ApJ...691..640R,2009Sci...325..171R,2011ApJ...740...15R}, where outflows dominate the flow structure at all depths.  

The moat flow as a distinct flow structure could not be left unnoticed by helioseismologists. \cite{2000JApA...21..339G} used an iterative deconvolution of the surface gravity ($f$) mode travel times and found that the properties of the moat flow averaged over the depth of $\sim1$~Mm are similar to those observed at the surface. Subsequent flow inversions around a cylindrically symmetric sunspot NOAA~9787 \citep{2009SSRv..144..249G} showed that the outflow from the spot is present to the depths of at least 4.5~Mm and gets stronger with depth. \cite{2011JPhCS.271a2002F} not only confirmed these conclusions by saying that the outflow region extends to depths at least 10~Mm, but also suggested that the outflow region had two components: a surface moat flow and a deeper part reaching the peak at depth around 5~Mm. 

So it would seem that the structure and the origin of the moat flow around sunspots is well understood. However, there is one piece of the puzzle missing: eventhough the moat flow is generally considered to be \emph{mostly} horizontal, it certainly must have some vertical component. Supergranular flows are also mostly horizontal and yet, there is an important weak (but measurable) vertical component with an upflow of around 4~\mps{} \citep{2010ApJ...725L..47D} in the centre of the cell and a downflow in the network lanes. The vertical component of the moat flow is badly addressed by observations. \cite{2013AA...551A.105L} concluded that the vertical components of both the Evershed and the surface moat flows are \emph{small} and that they even cannot address the sign of the vertical component. Due to the lack of observational evidence, the vertical component of the flow around sunspots is one of the targets of our study. It seems that currently only the state-of-the-art time--distance helioseismology is able to provide firm estimates of the weak vertical flows in the upper solar convection zone. 

\section{Linear inversions for time--distance helioseismology in a nutshell}

The investigated flows are measured using a time--distance helioseismology \citep{1993Natur.362..430D}. This method comprises a set of tools used to measure and analyse the travel times of solar waves travelling through the solar convection envelope. The propagation of the waves is affected by perturbed plasma parameters, which act as scatterers to the wave field. An important scatterer leaving a large imprint in \emph{difference} travel times (the difference of the measured travel time of waves travelling in  opposite directions) is a plasma streaming, which we want to study. 

The standard time--distance helioseismic pipeline consists of the following consecutive steps: first the spatio-temporal datacube is prepared using the tracking and mapping pipeline, this datacube is spatio-temporarily filtered to retain only waves of interest and subsequently the travel times are measured from cross-correlations of the filtered signal at two places. These travel times are finally inverted for flows assuming the linear relation between the flow vector and the measured travel time. 

In this study we analyse only the difference travel times measured using the surface gravity ($f$) mode of solar oscillations utilising the centre-to-annulus and centre-to-quadrant geometries \citep{1997SoPh..170...63D} with radii of the annuli 5 to 20 pixels with the step of 1 pixel and the pixel size of 1.46~Mm (henceforth ``travel times''). The travel times are fitted from the measured cross-correlation using \cite{2004ApJ...614..472G} linearised approach, the wave sensitivity kernels are computed using the Born approximation \citep{2007AN....328..228B} consistently with the travel-time measurements, and the travel-time noise covariance matrix is measured by fitting from the large set of travel-time maps \citep{2004ApJ...614..472G}. 

\begin{figure}
\includegraphics[width=\textwidth]{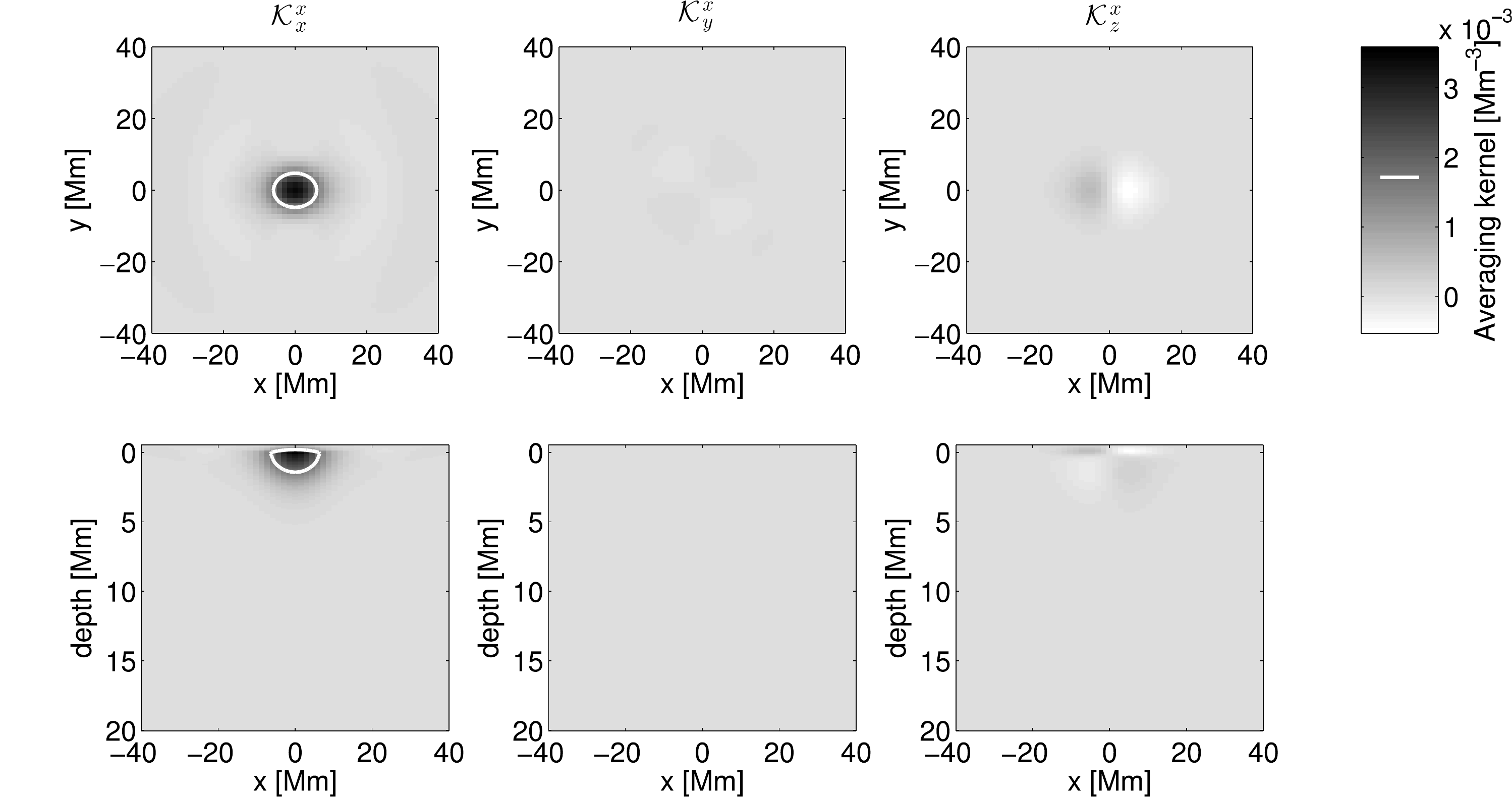}\\
\rule{\textwidth}{1pt}\\
\includegraphics[width=\textwidth]{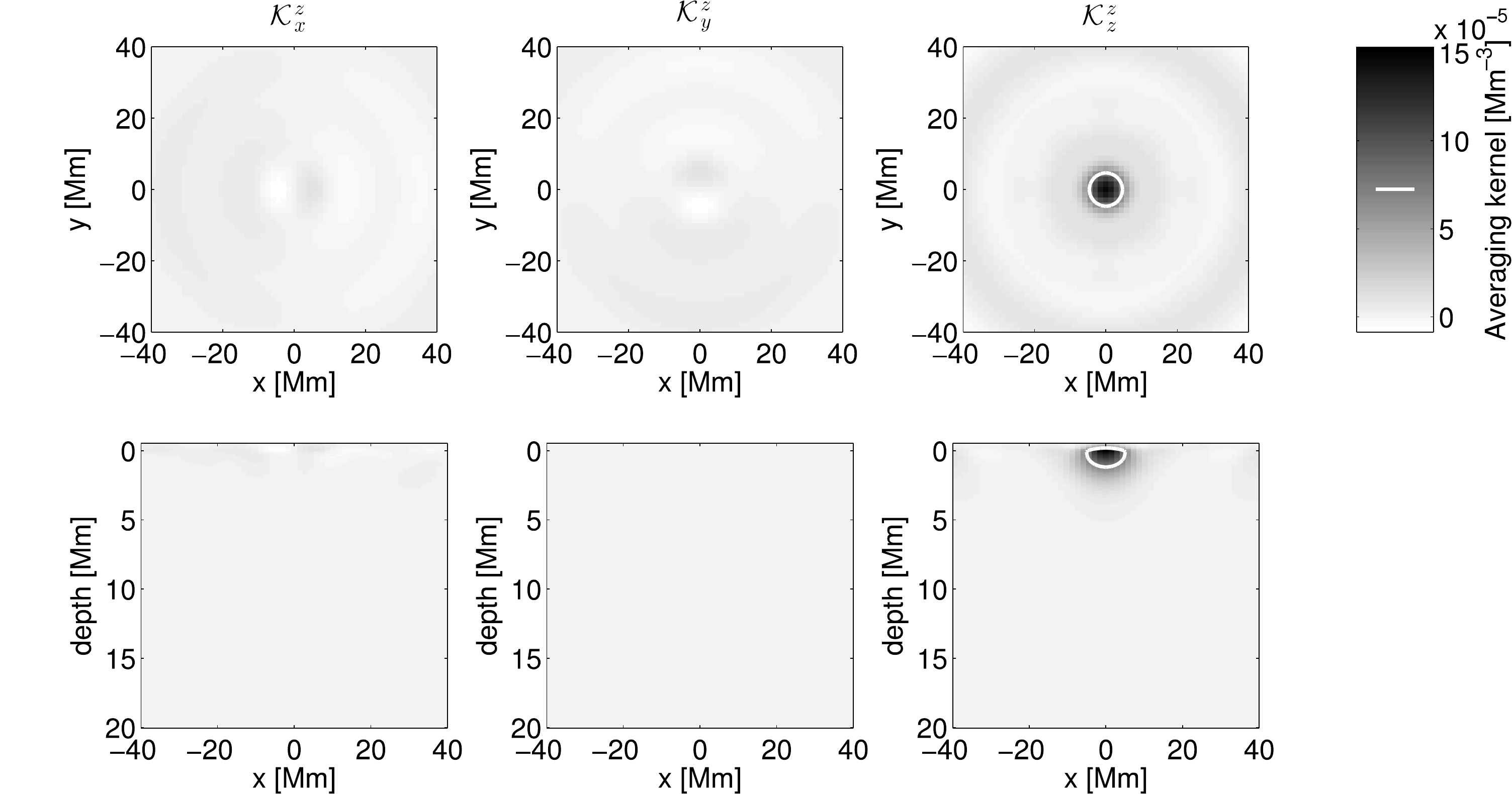}\\
\caption{Time--distance inversion averaging kernels. The upper set of panels represents the display of the kernel of horizontal components of the flow ($v_x$ in this case), while the lower set of panels is for the vertical flow component. The solid white line contours 50\% of the kernel maximum. }
\label{fig:akerns}
\end{figure}

The inversion is performed utilising Multichannel Optimally Localised Averaging approach \citep[MC-SOLA:][]{2012SoPh..276...19J} using a code validated using synthetic data \citep{Svanda2011}. The vector flow inversion differing only slightly was used by \cite{2013ApJ...771...32S}, where the whole time--distance pipeline was validated against the direct surface-flow measurement from granules tracking. Our flow estimates are thus representatives of the near sub-surface flow at depths of 0--3~Mm, with random-noise levels of 30~\mps{} for horizontal and 4~\mps{} for vertical components assuming travel-time averaging over 24 hours. The effective resolution of the flow map is set by the horizontal extent of the inversion averaging kernel (Fig.~\ref{fig:akerns}), which is also returned from the inversion. The horizontal full width at half maximum of the averaging kernel is 10 Mm. From Fig.~\ref{fig:akerns} one immediately notices two facts: the horizontal shape of the averaging kernel for horizontal flows components is slightly elliptical, while for the vertical component it is perfectly roundish, and the cross-talk contributions are negligible.

We would like to stress out that we are not comparing the inferences about the moat flow obtained by methods utilising the direct surface measurements with our time--distance results. That is because the time--distance inferences represent the real solar flow smoothed by the averaging kernel. This effect usually smears out any details in the flow, which exist on the scales smaller than the appropriate extent of the averaging kernel. From the same reason, our time--distance inferences do not represent surface plasma flow, but represent an average over the depth of 0--3~Mm, where the gravity centre of the averaging kernel lies at the depth of 1.0~Mm. One has to keep that on mind when discussing the time--distance inferences in the frame of other works and studies that speak about the purely surface inferences. 

\section{Ensemble averaging}

We processed 38 months of high-cadence (one frame each 45 seconds) full-disc Helioseismic and Magnetic Imager \citep[HMI:][]{2012SoPh..275..229S} Dopplergrams covering the period from May 1st 2010 to June 30th 2013. On each day we tracked the central-meridian region and mapped it using Postel projection with a pixel size of 1.46~Mm utilising a standard tracking tool\footnote{The tracking of Dopplergram datacubes was performed using the resources of Science Data Center for SDO at Max-Planck-Institut f\"ur Sonnensystemforschung, Germany. The filtering, travel-time measurement, inversions and the data analysis were then performed using the {\sc Sunquake} cluster of the Solar Department of Astronomical Institute of ASCR in Ond\v{r}ejov.}. The tracking and mapping resulted in a large set of datacubes that were filtered to retain only the signal of the $f$ mode and inverted for all three components of the flow independently. For each day we obtained one map for the vector flows in the central-meridian region, roughly between $\pm$70 degrees of latitude and $\pm$30 degrees in longitude from the disc centre. 

The magnetic field acts as an additional scatterer to the seismic waves, generally leading to a reduction of an acoustic power in the magnetised regions \citep[e.g.][]{1999ApJ...513L..79B}. To mitigate this problem, we normalised the measured cross-correlation at each point by its maximum value \citep[similarly to][]{2012SoPh..275..357C}. This approach allows to correct for waves absorption in regions occupied by a weak dispersed magnetic field. Such an approach does not improve the situation in the regions of the strong field, where also the physics of interaction of waves with magnetic field is not known well. Therefore, we do not consider the inferred flow in sunspots trustworthy \citep[for discussion see][]{2009SSRv..144..249G} and do not discuss these regions further. 

Due to the large noise level we chose not to study individual representatives of moats around sunspots and compare them to individual supergranular cells, but we rather proceed by using a statistical approach by forming the \emph{average representatives} of both features using the ensemble averaging approach. 

\subsection{Sunspots}
In order to make our analysis simpler, we chose to investigate the moat flows only around axially symmetrical sunspots of type H \citep{1990SoPh..125..251M}. We further put the following constraints on the spots belonging to the sample: the spot must be isolated with no other spot within 10 heliographic degrees distance, the spot at the time of observation should not be located farther than 20 heliographic degrees from the central meridian, and its latitudinal distance from the disc centre should be less than 30 degrees. The selection was done by hand in the first step, closely cooperating with NASA's www.solarmonitor.org, the accurate positions were then fine-tuned automatically from HMI intensitygrams by finding the gravity centres of the sunspots. 

Altogether we identified 104 spots fulfilling these constraints. The sizes of spots in the sample vary; the mean distance from the spot's centre to the outer penumbral boundary determined from corresponding HMI intensitygrams is 9.9$\pm$3.6~Mm. To account for the different sizes of sunspots in the ensemble, we normalised the inverted flow maps so that the outer penumbral boundary coincided for all spots. This normalisation was achieved by the interpolation of the flow maps onto a new coordinate grid, where the radial coordinate was added or subtracted a small correction so that the outer penumbral boundary lay at the distance of 10~Mm from the gravity centre of each spot. Such transformation conserves the distances in the radial direction but slightly distorts the distances in the tangential direction. It was shown by \cite{2007AA...472..277S} that the width of the moat depends only weakly on the size of the parental spot and given its average width (around 10~Mm), the distortions caused by our normalisation are negligible. The transformed maps of flows were averaged about the positions of gravity centres of respective HMI intensity images. When assuming that each sunspot flow map contains an independent realisation of the random noise component, the estimate for the error levels in each point is then 2.9~\mps{} for the horizontal and 0.4~\mps{} for the vertical component. These values are fully consistent with root-mean-square values of the quiet-Sun portions of the averaged flow map.



\subsection{Supergranules}

As a controlling set, we constructed a flow map of an average quiet-Sun supergranule. As a proxy for identification of supergranules, we used centre-to-annulus travel-time maps for distances 5--7 pixels (7--10~Mm), which were sensitive to a weak vertical flow and also to a divergence of the horizontal flow. Hence the supergranules were identified by searching the map for compact regions of large positive divergence (hence the negative travel time) surrounded by a region of negative divergence (hence positive travel time). The segmentation of individual supergranular cells was done using a watershed algorithm \citep{1992Watershed..433B}.

In a continuous space, the watershed algorithm recognizes individual basins belonging to regional minima $m_i$ of function $f(x)$ as sets of points having coordinates $x\in\mathbb{R}^2$ fulfilling the condition
\begin{equation}
f(m_i)+T_f(m_i,x)<f(m_j)+T_f(m_j,x)
\end{equation}
for all regional minima $m_j$, $j\neq i$, where $T_f(p,q)$ is the topographic distance of points $p$ and $q$, defined as
\begin{equation}
T_f(p,q)=\inf_\gamma\int_{\gamma}\id s\,\| f(\gamma(s))\|,
\end{equation}
where $\gamma(s)$ is a parametric curve connecting points $p$ and $q$. In the case of supergranules, the function $f$ is the travel-time value at coordinates $x$. In our case, we only consider local minima with travel-time value less than zero, thus excluding minima which are very unlikely to be a supergranular centre. By implementing this algorithm to a discrete set of points in the travel-time map we were able to assign a set of points to each selected regional minima (all minima with a positive plasma outflow).

The large number of travel-time maps allowed us to uniquely identify 222\,976 supergranular cells. All three-component flow maps obtained by the inversion were averaged about these supergranular cells. Again, assuming the independence of the random noise realisation, the random error levels are practically negligible (formally less than a few c\mps). 

\section{Comparison of flow patterns}

\begin{figure}
\includegraphics[width=\textwidth]{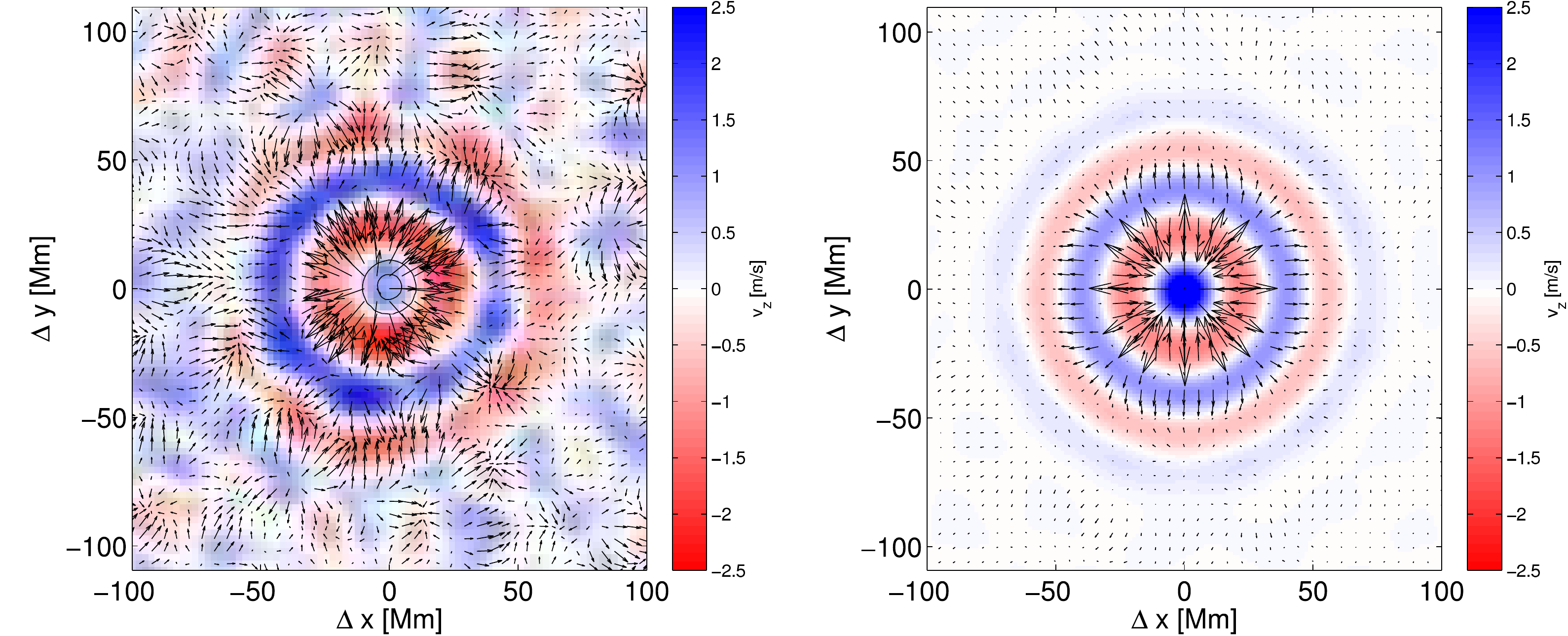}
\caption{Comparison of vector flow around an average H-type sunspot, marked by two concentric black circles, and an average supergranule. North is on the top and west on the right. The horizontal velocity component is depicted by arrows, the vertical one by colour coding.
Major differences seen are the east-west asymmetry of the horizontal moat flow around the sunspot and the difference in the magnitude of vertical velocity within the moat. The upflow ring at the distance of 38~Mm indicates the average location of neighbouring supergranules. The largest arrow indicates the horizontal flow of 400~\mps.}
\label{fig:flowmaps}
\end{figure}

\begin{figure}
\includegraphics[width=\textwidth]{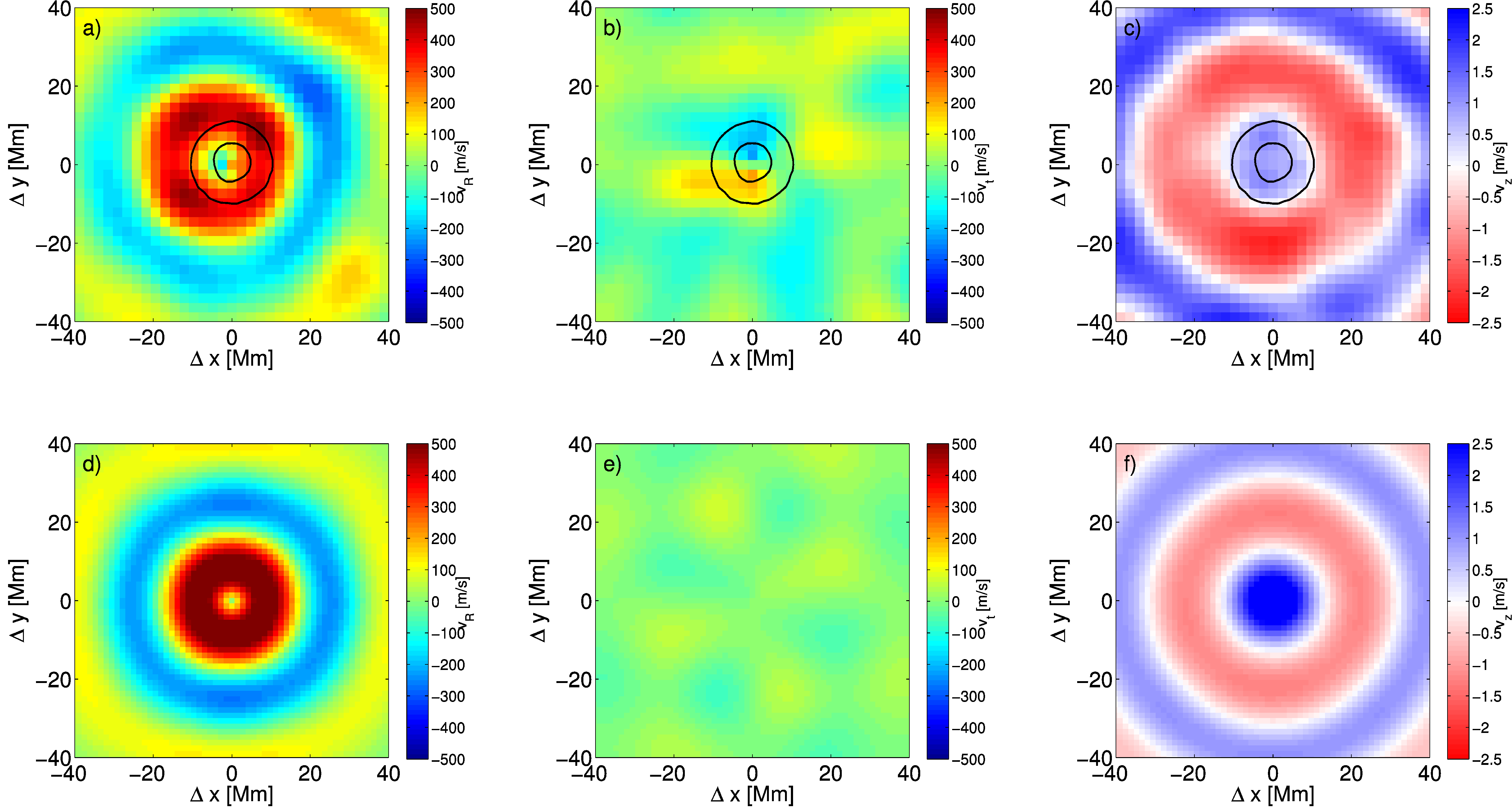}
\caption{Comparison of the flow components around an average H-type sunspot (two concentric black circles) and an average supergranule for reference. Maps of radial (a), tangential (b), and vertical (c) flow components around the average H-type spot and corresponding maps of radial (d), tangential (e), and vertical (f) flow components in the average supergranule show a clear distortion of the horizontal moat outflow due to the westward proper motion of the sunspot with respect to the local frame of rest (a). The moat shows up clearly as a downflow region (c). The tangential component is positive in the counter-clockwise direction. The sector-like structure of the tangential component of the horizontal flow in the average supergranule (e) is an artefact caused by the slightly elliptical shape of the inversion averaging kernel in the horizontal domain.}
\label{fig:components}
\end{figure}

\begin{figure}
\includegraphics[width=\textwidth]{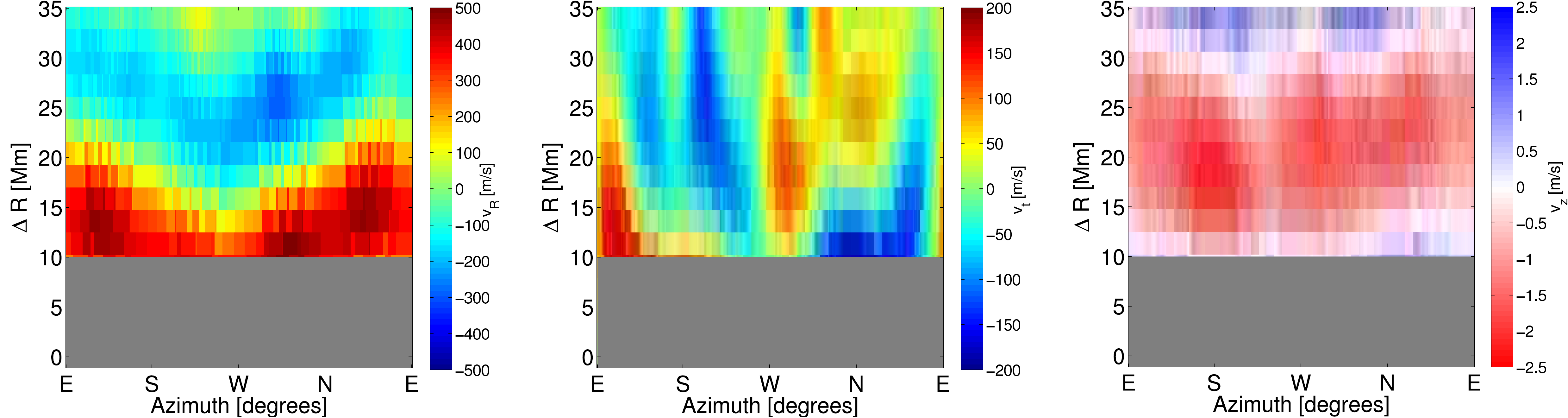}
\caption{Polar-coordinates plots of Figs.~\ref{fig:components}a-c to show better the asymmetry of the horizontal components of the moat outflow. Contrary to the radial and tangential (positive = counter-clockwise) components, the vertical component does not show any remarkable asymmetry, except a slightly enhanced downflow to the south and to the north of the spot. The shaded region indicates the strong-field regime, where the time--distance inferences are not trustworthy. }
\label{fig:asymmetries}
\end{figure}

\begin{figure}
\includegraphics[width=0.5\textwidth]{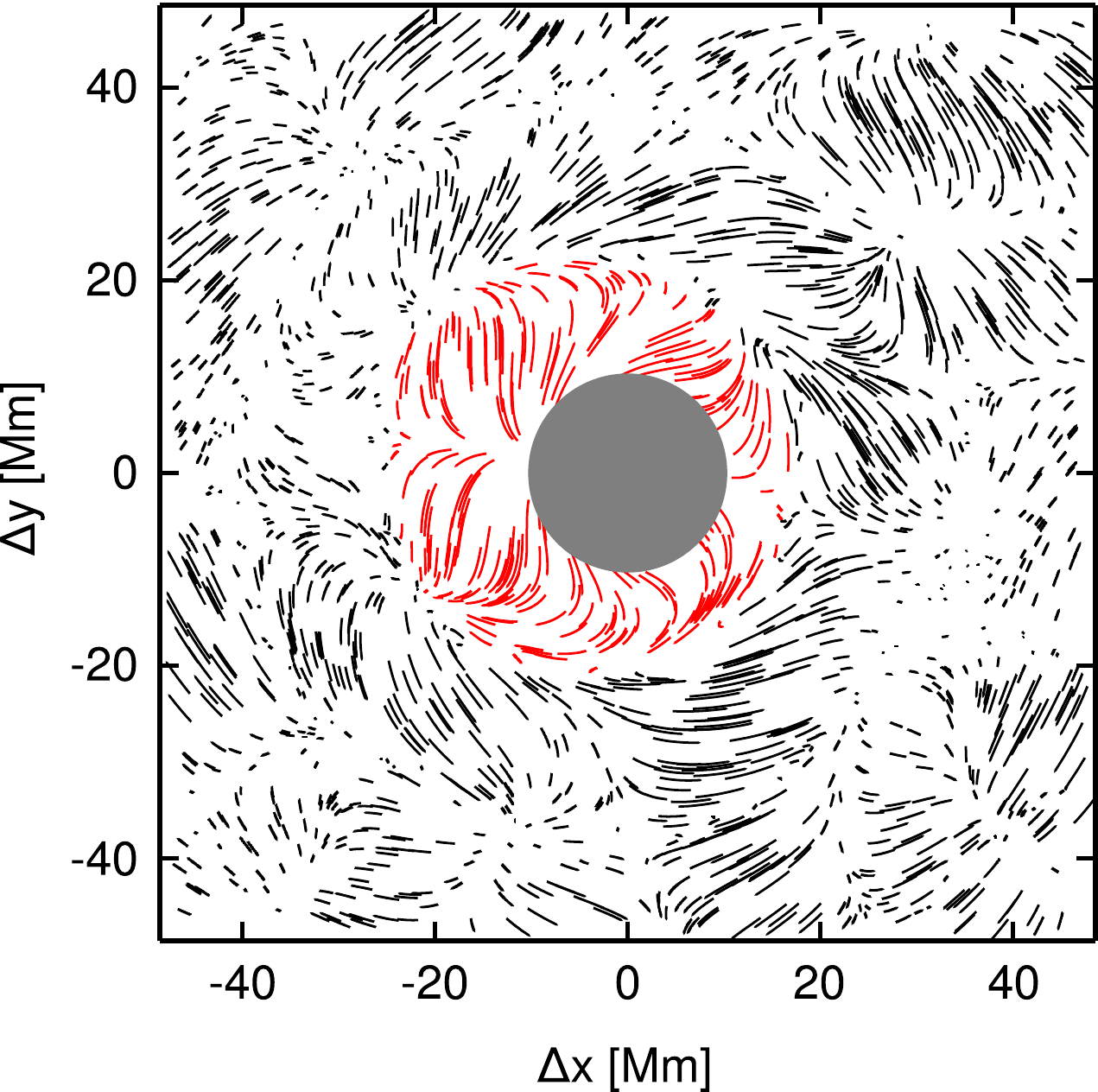}
\caption{A sketch of the horizontal streaming around an average H-type sunspot in the local frame of rest. In this plot, the tangential component of the horizontal flow component was amplified by a factor of 5 to illustrate the effect of the outflow deflection by the proper motion of the sunspot in the positive $\Delta x$ direction. The grey circle represents the location of the sunspot, the streamlines within the moat are coloured in red. }
\label{fig:streams}
\end{figure}

\begin{figure}
\includegraphics[width=\textwidth]{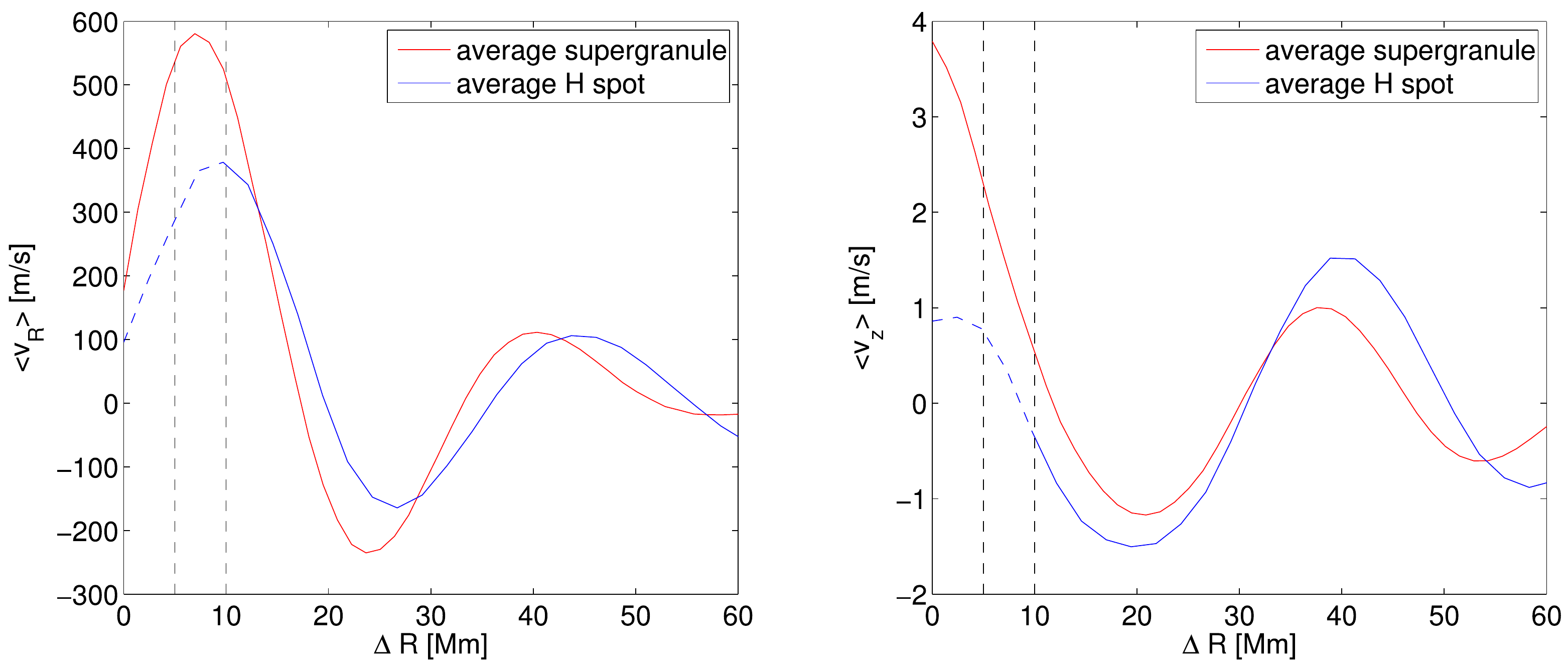}
\caption{Comparison of azimuthally averaged radial profiles of horizontal (left) and vertical (right) velocity components around an average H-type sunspot and in an average supergranule. The dashed segments of the graphs are those not considered trustworthy due to the presence of a strong magnetic field. Vertical dashed lines correspond to the radii of umbra and penumbra.}
\label{fig:profiles}
\end{figure}

\begin{figure}
\includegraphics[width=\textwidth]{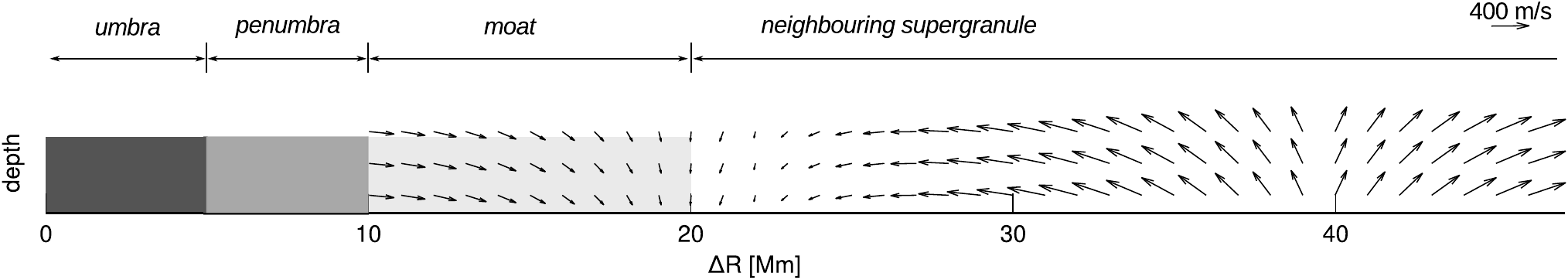}
\caption{A simplified model of the flow system around a H-type sunspot, including a neighbouring supergranule. In this plot, the magnitude of the vertical component of the velocity vectors is amplified by a factor of 100.}
\label{fig:model}
\end{figure}

From the comparison of statistically significant samples it turned out that the moat flows around symmetric H-type spots and the outflows within the supergranular cells
are  similar (see Fig.~\ref{fig:flowmaps}). There are, however, two principal differences.
\begin{enumerate}

\item While the outflow region within the average supergranular cell is very symmetric about the centre of the cell (that claim is true even when lesser number of supergranules, comparable to the number of the sunspots used here, is averaged), the moat outflow region displays a clear asymmetry in the east-west direction (Fig.~\ref{fig:components}). Such an asymmetry was already found by \cite{2007AA...472..277S} -- see Section 1. They compared average areas of 20$^\circ$ wide sectors of moats around well-developed spots. The sector directed to the east had the area larger by a factor of 4--5 than that directed to the west. Our data are consistent with this finding: the moat outflow extends from the penumbral boundary by 16~Mm to the east, which is about twice the extent of 7~Mm to the west (Fig.~\ref{fig:asymmetries}). According to \cite{2007AA...472..277S}, the moat outflow is distorted, due to a viscosity of gas, by proper motion of sunspots to the west. The mass conservation in the distorted radial outflow and a nearly symmetrical downflow is kept due to the asymmetry in the tangential component of the flow. The radial outflow is redirected (by the sunspot's proper motion) around the spot, first to the north and south and then eastward. This is seen in both the radial and tangential components of the flow (for illustration, see Fig.~\ref{fig:streams}). A slight asymmetry is also seen in the vertical component of the moat flow, however it is not as pronounced as in the case of the radial and tangential components of the horizontal flow. Note that all the discussed figures are plotted in the local frame of rest, in which the sunspot drifts westwards and the visualisation of the flow field differs from the natural comoving frame. Unfortunately, it is not possible to easily transform between the two frames, as the drifting speed is unknown and cannot be determined reliably, because the only observable that characterises the environment at the studied depth is the flow field. A very rough estimate can be made by decomposing the flow field into the symmetrical and non-symmetrical parts, where the non-symmetrical part at the sunspot location represents the drifting speed. Using such procedure we estimate the drifting speed to be 120~\mps. Eventhough such number is in an excellent agreement with the drifting speed for old large sunspots found by \cite{2007AA...472..277S}, we do not consider our determination very reliable given the assumptions made.

\item
Azimuthally averaged radial profiles of horizontal and vertical velocity components (Fig.~\ref{fig:profiles}) and a cartoon displaying a simplified model of velocity vectors in the moat and the neighbouring supergranule (Fig.~\ref{fig:model}) show that within the supergranular cell, there is an upflow near its centre, which turns into a downflow at some 60\% cell radius from the cell centre.
The moat is a \emph{purely} downflow region with a slight asymmetry, extending from the penumbral boundary by about 12 Mm, where it is adjacent to downflows at the borders of neighbouring supergranules. The maximum azimuthally averaged downflow speed in the moat is 1.5 \mps, larger by a factor of 1.3 than that at the supergranular border. The out- and downflow in the moat region should be compensated by an upflow in the region of the sunspot, which is not detectable by the present helioseismic methods and may serve as a possible mechanism enhancing the moat downflow. 
\end{enumerate}

The average distance of the surrounding supergranules from the spot centre is 40~Mm, which is only slightly larger than the average distance between centres of neighbouring supergranular cells (38~Mm; see Fig.~\ref{fig:flowmaps}). This would favour the hypothesis that an isolated medium-sized symmetrical sunspot and the flow system around it (the moat flow) acts on average as a larger supergranular cell \citep{1965IAUS...22..192B}. Formation of the moat flow replacing the ordinary supergranular flow was also described in some of the early models of sunspots formation \citep[e.g.][]{1976ApSS..40...73P,1976ApSS..41...79P}.

\subsection{Mass flow rates}
We estimated the mass flow rates within the average supergranular cell and within the moat region of the average H-type spot. We evaluated the radial cumulative mass flow rate
\begin{equation}
\dot{m}(R)=\int\limits_{0}^R \id R\,R\,\rho \,v_z,
\end{equation}
where $\rho$ is the density estimated as an average of the model-S density \citep{1996Sci...272.1286C} weighted by the inversion averaging kernel,
$v_z$ the vertical velocity component,
and $R$ is the radial coordinate. We assume the density to be constant in the horizontal domain. 

In the case of the average supergranule, $\dot{m}(R)$ reaches its maximum of $\sim 5\times 10^5$~\kgs{} at the distance of $\sim12$~Mm from the cell centre and then vanishes at the distance 19~Mm, which corresponds to the supergranule boundary. From this point of view, the continuity equation holds within the supergranular cell under the assumptions. Such result may be interpreted also in a different way: the plasma density within the supergranular cell does not vary much in the horizontal domain.

In the case of the average H-type sunspot, when ignoring the strong-field regions and neglecting the east-west asymmetry, we obtain the total cumulative mass flow rate of $\dot{m}(R)\sim -10\times10^5$~\kgs. Since we ignored both the sunspot umbra and penumbra, this number must be considered as a lower limit. Hence in the shallow sub-surface around the H-type sunspot, at least twice more mass sinks than in the average supergranular cell. This downflow must be compensated by an upflow somewhere in the magnetised region of sunspot. The structure and amplitude of such upflow cannot be determined using present helioseismic techniques. 
When assuming that a homogeneous upflow takes a shape of an annulus with a width of 500~km and a radius of 5~Mm, approximately under the boundary of the umbra, then the lower limit of such homogeneous upflow would be 20~\mps. Without knowing the real structure of the flow under the umbra we cannot do much more at this stage. 

\section{Concluding remarks}
Compared to surface velocity observations, speeds obtained by helioseismic methods are usually small. Our maximum horizontal velocity of the moat outflow, 400~\mps, is nearly identical with the result of \cite{2007AA...472..277S}. However, the magnitude of the Evershed flow is by an order larger and vertical velocities observed on the solar surface in and around penumbrae are larger by 2--3 orders than the subsurface downflow in the moat. 
(i) Current helioseismic methods are restricted to non-magnetic and weak-field regions, so that we miss the information from the penumbra. (ii) All extremely fast flows on the surface, including the Evershed flow, are localised to small areas and vary in time, so that they are smoothed by spatial and temporal averaging in helioseismic measurements. 

The subsurface moat flow surrounding a symmetric H-type sunspot, obtained as an average of 104 sunspots, is in principle very similar to a flow system in an average supergranule (222\,976 cells averaged).
The average size of moat regions is comparable with the average size of supergranules, so that the moat flows around H-type sunspots seem to replace the ordinary supergranular flows.
However, the flows in the moat are asymmetrical. We expect that the westward proper motion of sunspots with respect to the local frame of rest distorts the horizontal outflow in the moat and redirect it partially back around the sunspot. This confirms the asymmetry detected previously in horizontal motions of granules by \cite{2007AA...472..277S}.

Thanks to the improved formulation of the MC-SOLA time--distance inversion, we were able, for the first time, to study properly the vertical component of velocity, in which moats differ from supergranules. The whole moat is a downflow region with the flow amplitude (1.5~\mps) larger than the downflows at the edges of supergranules, measured in our reference sample. Our conservative estimate shows that the mass submerging in the considered part of the moat is twice the mass circulating in the near-surface layers of the average supergranule. Hence, at least the same amount of mass must emerge close around the sunspot flux tube.
It should be possible to confront our measurements with the state-of-the-art numerical models of sunspots. 

\acknowledgements This work was supported by the Czech Science Foundation (grants 14-04338S and P209/12/P568). The data were provided by the HMI consortium via the resources at Max-Planck-Institut f\"ur Sonnensystemforschung, Germany, which is funded by the German Aerospace Center (DLR). Tato pr\'ace vznikla s podporou na dlouhodob\'y koncep\v{c}n\'\i{} rozvoj v\'yzkumn\'e organizace (RVO:67985815) a v\'yzkumn\'eho z\'am\v{e}ru MSM0021620860. 


\end{document}